\documentclass{article}

\usepackage{microtype}

\usepackage{deluxetable}
\makeatletter
\def\@to{to}
\makeatother

\usepackage{graphicx}
\usepackage{subfigure}
\usepackage{booktabs} % for professional tables
\usepackage{float}

\usepackage{hyperref}

\usepackage{float}

\usepackage[accepted]{icml2023}

\usepackage{amsmath}
\usepackage{amssymb}
\usepackage{mathtools}
\usepackage{amsthm}
\usepackage{CJKutf8}
\usepackage{appendix}
\usepackage[most]{tcolorbox}
\usepackage{xcolor}

\usepackage[capitalize,noabbrev]{cleveref}

\theoremstyle{plain}

\theoremstyle{definition}

\theoremstyle{remark}

\usepackage[textsize=tiny]{todonotes}

\icmltitlerunning{Conditional Normalizing Flows as a Gyrochronological Framework}

\begin{document}

\twocolumn[
\icmltitle{A Novel Application of Conditional Normalizing Flows: \\ Stellar Age Inference with Gyrochronology}




\begin{icmlauthorlist}
\icmlauthor{Phil Van-Lane}{daddaa}
\icmlauthor{Joshua S. Speagle \begin{CJK*}{UTF8}{gbsn}(沈佳士)\end{CJK*}}{stats,daddaa,ds,dunlap}
\icmlauthor{Stephanie Douglas}{lafayette}
\end{icmlauthorlist}

\icmlaffiliation{daddaa}{Department of Astronomy \& Astrophysics, University of Toronto, Toronto, ON, Canada}
\icmlaffiliation{stats}{Department of Statistical Sciences, University of Toronto, Toronto, ON, Canada}
\icmlaffiliation{lafayette}{Department of Physics, Lafayette College, Easton, PA, United States}
\icmlaffiliation{dunlap}{Dunlap Institute of Astronomy \& Astrophysics, University of Toronto, Toronto, ON, Canada}
\icmlaffiliation{ds}{Data Sciences Institute, University of Toronto, Toronto, ON, Canada}

\icmlcorrespondingauthor{Phil Van-Lane}{phil.vanlane@mail.utoronto.ca}

\icmlkeywords{Machine Learning, ICML}
\vskip 0.3in
]



\printAffiliationsAndNotice{}
% leave blank if no need to mention equal contribution
\begin{abstract}
Stellar ages are critical building blocks of evolutionary models, but challenging to measure for low mass main sequence stars. An unexplored solution in this regime is the application of probabilistic machine learning methods to gyrochronology, a stellar dating technique that is uniquely well suited for these stars. While accurate \textit{analytical} gyrochronological models have proven challenging to develop, here we apply conditional normalizing flows to photometric data from open star clusters, and demonstrate that a data-driven approach can constrain gyrochronological ages with a precision comparable to other standard techniques. We evaluate the flow results in the context of a Bayesian framework, and show that our inferred ages recover literature values well. This work demonstrates the potential of a probabilistic data-driven solution to widen the applicability of gyrochronological stellar dating.
\end{abstract}

\section{Introduction} \label{sec: intro}

Stellar evolution and lifecycle models are critically important to astronomy on a wide range of scales, but require properly constrained age data. Since its conception \cite{1964ApJ...140..544D, 1969ApJ...158..685S}, isochrone fitting has been the most ubiquitous technique for dating star clusters. Isochrones are population models for stars of the same age but a variety of initial masses, so this technique performs poorly in the regime of low mass main sequence (MS) stars due to their degeneracy with age in the colour-luminosity parameter space. This is particularly true in the case of non-coeval field stars with weakly constrained cluster membership, where the presence of multiple populations may hinder the identification of isochrones. Other stellar dating methods such as asteroseismology \cite{2007A&ARv..14..217C, 2022ARA&A..60...31K}, Lithium abundances \cite{2014EAS....65..289J, 2017A&A...602A..63B, 2019AJ....158..163D} and x-ray luminosity \cite{2017A&A...602A..63B,2019AJ....158..163D} struggle in this regime as well. One technique that avoids such issues is gyrochronology, which relies on the concept of stellar spin-down to infer age using measurements of rotation period and MS location (ie. mass, proxied with an observable such as colour). This makes gyrochronology uniquely well-suited for dating low mass MS stars, especially in the current era of abundant stellar rotation data from missions such as \textit{Kepler} and \textit{Transiting Exoplanet Survey Satellite (TESS)}.

Beginning with several foundational papers by Barnes \yrcite{2003ApJ...586..464B, 2007ApJ...669.1167B, 2010ApJ...722..222B}, most gyrochronological studies have leveraged empirical models calibrated on stellar population ages that were benchmarked using other methodologies \cite{2019AJ....158..173A, 2019AJ....158...77C, 2022arXiv221109822B, 2022AJ....164..137K}. Open star clusters have often been used for validation in such analyses; these are gravitationally-bound groups of tens to hundreds of stars formed concurrently, and are useful gyrochronological calibrators due to their effectiveness with isochrone fitting. Other studies have relied on physical models of stellar spin-down rates to predict rotational evolution \cite{2016Natur.529..181V, 2021ApJ...912...65G}. The reliability challenges presented by these techniques \cite{2015MNRAS.450.1787A, 2016Natur.529..181V, 2022arXiv221001137S}, especially in stars older than $\approx$ 1-2 Gyr, emphasize the difficulty in fitting empirical models to observations, and highlight the need for a data-driven approach. To the best of our knowledge, a gyrochronological model leveraging machine learning has not yet been implemented, and this work builds towards filling that gap by developing a probabilistic framework designed to infer stellar ages from real observational data.

\begin{figure*}[ht]
% \vskip 0.2in
\begin{centering}
\centerline{\includegraphics[width=450pt]{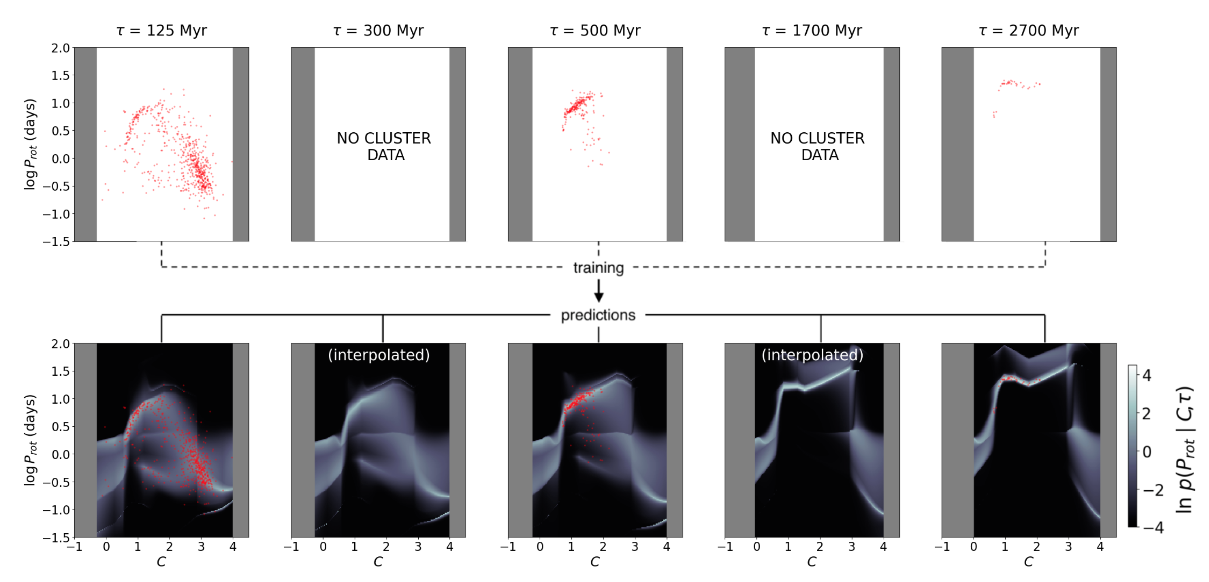}}
\caption{This figure shows subsets of our \textit{(i)} observational data represented by solid red circles, and \textit{(ii)} normalizing flow results represented by colour maps, both at 5 discreet sample ages. For ages aligned with our open clusters ($\tau$ = 125, 500, and 2,700 Myr), we show results from our cross-validation exercise (see \S\ref{sec: results}) in the bottom row. For ages that we do not have observational data for ($\tau$ = 300 and 1,700 Myr), the results are generated from a NF trained on all observations.}
\label{fig: nf_interp}
\end{centering}
% \vskip -0.2in
\end{figure*}

\section{Methods} \label{sec: methods}

\subsection{Data} \label{subsec: data}

The observational data used in this work consist of rotation period ($P_{rot}$) and colour (standardized to the Gaia de-reddened photometric magnitudes $(G_{BP}\!-\!G_{RP})_0$ but denoted as $C$ going forward for brevity) for 2,878 total stars from 8 different open clusters. Our data were curated from two different sources \cite{2020ApJ...904..140C, 2021ApJS..257...46G}, utilizing only the subset that each considered to be "probable" cluster members. De-reddening corrections have already been applied by Curtis \yrcite{2020ApJ...904..140C}, but were calculated by us using the colour data presented by Godoy-Rivera \yrcite{2021ApJS..257...46G}. These papers also provide age estimates of the open clusters, which we refer to throughout this work as the "literature" ages $\tau_{lit}$, and use as the basis of evaluation for our results (see Appendix \ref{sec: app_clustes} for a detailed breakdown).

\begin{figure*}[!ht]
\vskip 0.2in
\begin{center}
\centerline{\includegraphics[width=430pt]{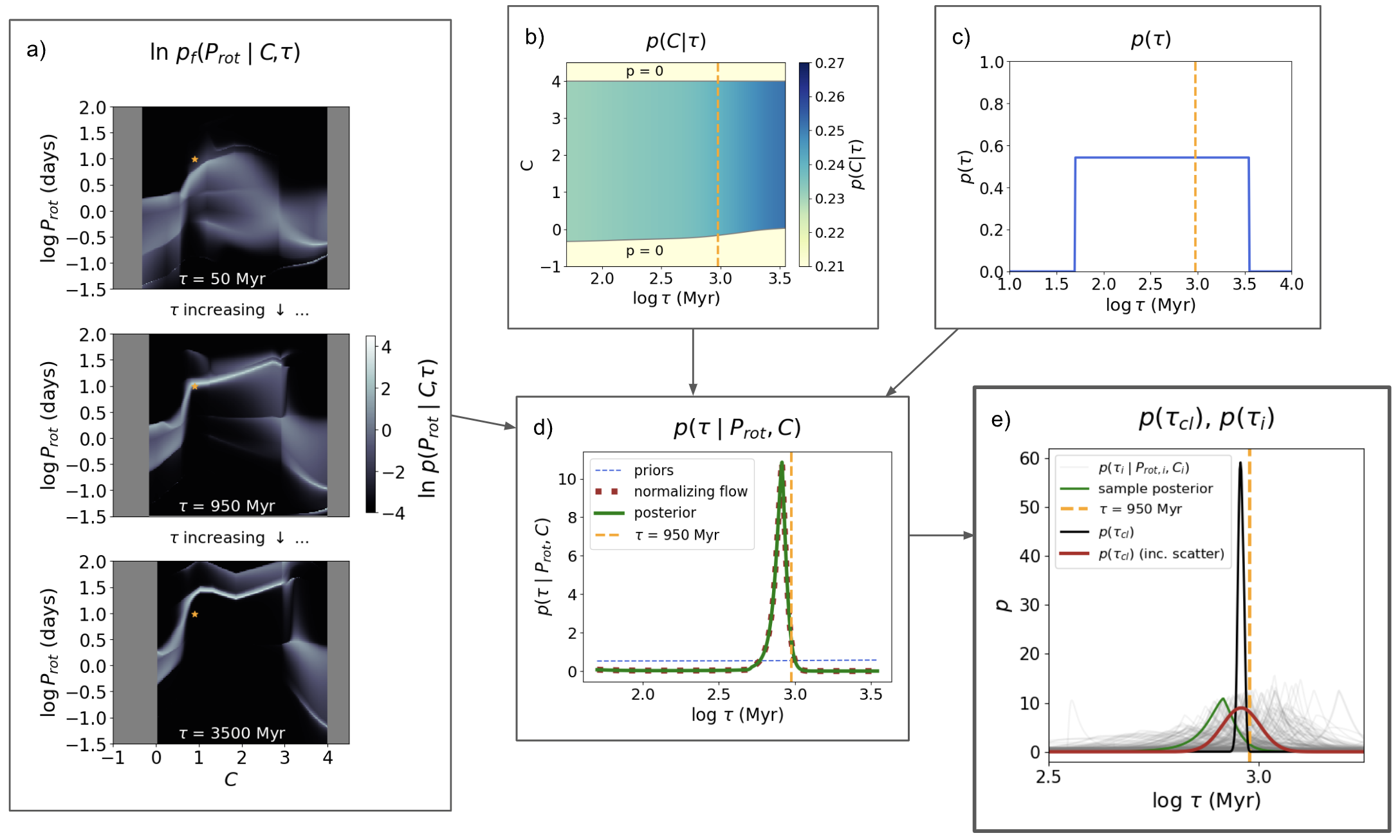}}
\caption{NF results, priors and posterior age distribution for a sample star from the NGC 6811 cluster ($P_{rot}$ = 9.75 d, $C$ = 0.910), and $p(\tau_{cl})$ for NGC 6811. Panel \textbf{a)} illustrates the NF-generated $p(P_{rot}\;|\;\tau\!,\!C$); the star's observed properties are indicated by the orange point. Panel \textbf{b)} shows the behaviour of the $p(C|\tau)$ across ages. Panel \textbf{c)} illustrates the uniform age prior. Panel \textbf{d)} demonstrates the calculation of the posterior age distribution for this sample star as described in eq. \ref{eq: bayes_with_var_expanded}. Panels \textbf{b)}, \textbf{c)}, \textbf{d)}, and \textbf{e)} all show the sample stellar age from literature as a dashed orange line. Panel \textbf{e)} shows $p(\tau_{cl})$ for NGC 6811 as calculated using eq. \ref{eq: cluster_post}; the black line illustrates the distribution using only statistical uncertainties while the red line represents that posterior convolved with the intrinsic scatter that was generated with our cross-validation tests. Individual stellar posteriors are also shown in translucent grey; the green line corresponds to the sample star's posterior.}
\label{fig: post_flow}
\end{center}
\vskip -0.2in
\end{figure*}

\subsection{Rotation period inference with normalizing flows} \label{subsec: nf}

The challenge motivating this work is predicting the evolution of a star's rotation period from its age ($\tau$) and colour. We describe this as the conditional probability: $p(P_{rot}\;|\;C,\tau)$, which fits into our Bayesian gyrochronological framework as formulated in eq. \ref{eq: bayes_with_var_expanded}. This equation is expanded from Bayes' theorem via a decomposition of our observational likelihood term $p(P_{rot}, {C}\;|\;\tau)$ into the product of $p(P_{rot}\;|\;C,\tau)$ and $p(C\;|\;\tau)$, and is expressed in terms of our posterior probability for the age of a star $p(\tau\;|\;P_{rot}\!,\!C)$:

\begin{equation} \label{eq: bayes_with_var_expanded}
    p(\tau\;|\;P_{rot}, C) = \frac{p(P_{rot}\;|\; {C},\tau) \cdot p(C\;|\;\tau) \cdot p(\tau)}{p(P_{rot},C)}
\end{equation}

To calculate $p(P_{rot}\;|\;C,\tau)$, we implement a normalizing flow (NF), taking advantage of this method's unique ability to model conditional probability densities across a continuous parameter space. We train our NF to learn the conditional $p_f(P_{rot}\;|\;C\!,\!\tau)$ distribution across our observational data ($p_f$ now representing the probability calculated by the flow). Our NF is built using the \texttt{pyro.ai}\footnote{\url{https://github.com/pyro-ppl/pyro}} framework \cite{bingham2018pyro, phan2019composable}, a python-based package that extends the \texttt{PyTorch} deep learning library, and we implement the Adam optimizer for our training \cite{2014arXiv1412.6980K}. We use a series of 4 layers to train our flow: in order of implementation they are \textit{(i)} linear spline \cite{2019arXiv190604032D, 2020arXiv200105168D}, \textit{(ii)} matrix exponential \cite{2018arXiv180205957M, 2016arXiv160207868S, 2020arXiv200601910H}, \textit{(iii)} householder \cite{2016arXiv161109630T}, and \textit{(iv)} affine autoregressive \cite{2015arXiv150203509G, 2015arXiv150505770J, 2016arXiv160604934K} transformation layers. These were chosen based on empirical performance during our model evaluation stage. Each training run includes 150 000 steps with a learning rate of $5e\!-\!4$. Our loss function $L$ minimizes the negative log probability of the conditional $p(P_{rot}\;|\;C\!,\!\tau)$ distribution. We further decompose this probability into two components: the $p_f$ computed by our NF and a background probability $p_b$ that represents a uniform distribution across rotation period that we expect to see in non-cluster field stars:

\begin{equation} \label{eq: bg}
    p_{b}(P_{rot}) \sim 
    \begin{cases}
    \frac{1}{P_{rot}^{max} - P_{rot}^{min}} & P_{rot}^{min} < P_{rot} < P_{rot}^{max}\\
    0 & otherwise\\
    \end{cases}
\end{equation}

$P_{rot}^{min}$ and $P_{rot}^{max}$ are the limits of $log$ $P_{rot}$ (in days) that we evaluate our loss over; in this case -1.5 and 2.0 respectively. With this additional layer of nuance, and $w_f$ applied as a weighting of our NF probability against our background probability, our loss function becomes (given data for each $i^{th}$ star out of $n$ total stars):

\begin{equation} \label{eq: loss}
    \begin{split}
    L = \Sigma_{i=1}^n [-\ln(w_{f} \cdot p_{f}(P_{rot,i}\;|\;C_i,\tau_i)\\+
    (1-w_{f}) \cdot p_{b}(P_{rot,i}))]
    \end{split}
\end{equation}

The inclusion of $p_b$ is important to minimize training artifacts when considering the inherent uncertainty in observational data, as there is always natural scatter and potential for field star contamination. We apply $w_{f} = 0.90$ in this work, as this value is the criteria for "probable" cluster membership according to Godoy-Rivera et al. \yrcite{2021ApJS..257...46G}. However, using a dynamic or fitted value for this weighting is a future optimization that we intend to apply.

\section{Results} \label{sec: results}

Figure \ref{fig: nf_interp} displays our NF results at a subset of discreet ages. The top row of subplots shows only observational data (where available), whereas the bottom row overlays the observations atop the NF-generated $p_f(P_{rot}\;|\;{C},\tau)$ probability grid at each age value. The second and fourth panels in the bottom row are examples of NF results at ages for which we do not use any observational data in this work, and hence cannot validate directly. The behaviour of the flow at these ages is important when considering the challenge of heavily clustered data points in age space; only eight different age values are represented across all 2,878 data points, with significant distance in age space between many of the clusters. Despite these large gaps in training data, we see that the NF interpolation does not behave erratically. Another important note is that the seemingly high probabilities in regions of sparse data at the extremes of the colour spectrum are due to the conditional nature of our probability distributions; since each subplot in the bottom row represents $p_{f}(P_{rot}\;|\;C\!,\!\tau)$ at a specific age slice, the probability distribution along any vertical line through a subplot will integrate to 1. Hence, these "wings" represent the $p_f(P_{rot})$ distribution that we would expect to see given stars at those ages and colours.  This illustrates the importance of applying a \textit{conditional} NF to mitigate selection effects, which otherwise could bias our NF in colour space.

To evaluate performance within the context of gyrochronological age estimation, we apply the NF results to eq. \ref{eq: bayes_with_var_expanded} to calculate posterior age distributions, and compare those to $\tau_{lit}$ from our data sources \cite{2020ApJ...904..140C, 2021ApJS..257...46G}. We perform this analysis using leave-one-out cross validation (LOOCV), training 8 different NFs and for each one excluding observational data from a single cluster (which is then subsequently used as the validation data set for the NF). Additionally, to apply our NF results to equation \ref{eq: bayes_with_var_expanded}, we require  \textit{(i)} $p(\tau)$ and \textit{(ii)} $p(C|\tau)$. These are calculated based on a uniform distribution of $\tau$ (in Myr) of:

\begin{equation} \label{eq: age_prior}
    \tau \sim U(50,3500)
\end{equation}

and a uniform distribution of C at any given $\tau$ of:

\begin{equation} \label{eq: colour_dist_prior}
    \begin{aligned}
        (C\;|\;\tau) \sim U(C_{iso(\tau)}^{min},C_{obs}^{max})
    \end{aligned}
\end{equation}

We utilize isochrones generated from the \texttt{brutus}\footnote{\url{https://zenodo.org/record/3840241}} package to constrain the lower limit on the colour values for a star at any given age: $C^{min}_{iso(\tau)}$. The upper limit $C^{max}_{obs}$ is set to a constant of the largest (reddest) colour value present in the entire observational data set. These priors are relatively uninformative compared to stellar models available, which is an intentional choice that mitigates selection effects which may be generated by observational bias and inconsistent instrument behaviour. After computing all individual stellar age posteriors, we proceed to calculate overall cluster posteriors $p(\tau_{cl})$. A more thoughtful approach to generating these will be applied in future work, however in this initial phase we calculate $p(\tau_{cl})$ for each cluster as a naive product of the individual stellar posteriors for all $n_{cl}$ stars in the cluster:

\begin{equation} \label{eq: cluster_post}
    p(\tau_{cl}) = \prod_{i = 1}^{n_{cl}} p(\tau_{i}\;|\;P_{rot,i}, C_i)
\end{equation}

Once we have these cluster age posteriors (described with means $\tau_{inf}$ and statistical uncertainties $\sigma_{\tau_{inf}}$) inferred from our NF results, we additionally apply Markov Chain Monte Carlo (MCMC) using the \texttt{emcee}\footnote{\url{https://github.com/dfm/emcee}} package \cite{2013PASP..125..306F} to fit our data to:

\begin{equation} \label{eq: mcmc}
    \tau_{inf} \sim N(\theta \cdot \tau_{lit}, \sigma_{\tau_{inf}^2} + s^2)
\end{equation}

which includes an intrinsic scatter term $s$.

Figure \ref{fig: post_flow} illustrates visually how eq. \ref{eq: bayes_with_var_expanded} is applied to generate posterior age distributions for single stars (\textbf{a} + \textbf{b} + \textbf{c} $\rightarrow$ \textbf{d}), how eq. \ref{eq: cluster_post} is applied to generate the cluster posteriors (\textbf{d} $\rightarrow$ \textbf{e}), and the effect of our fitted intrinsic scatter from eq. \ref{eq: mcmc} on the shape of $p(\tau_{cl})$ (\textbf{e}). It is worth noting that the denominator $p(P_{rot},C)$ of eq. \ref{eq: bayes_with_var_expanded} is constant across our model, so can be ignored given that we normalize the results. As expected, panel \textbf{d)} demonstrates that since our $p(\tau)$  and $p(C\;|\;\tau)$ priors are nearly uniform (see eqs. \ref{eq: age_prior} and \ref{eq: colour_dist_prior}), the posterior is NF-dominated. It is apparent in panel \textbf{e)} that for this cluster, the posterior mean $\tau_{inf}$ is quite close to the nominal age of 950 Myr. It is also clear that including our intrinsic scatter term produces a much more reasonably constrained $p(\tau_{cl})$ than the over-precise distributions generated from only considering the statistical uncertainties (see Appendix \ref{sec: app_all_loocv_res} for individual cluster results).

\begin{figure}[!ht]
\begin{center}
\centerline{\includegraphics[width=200pt]{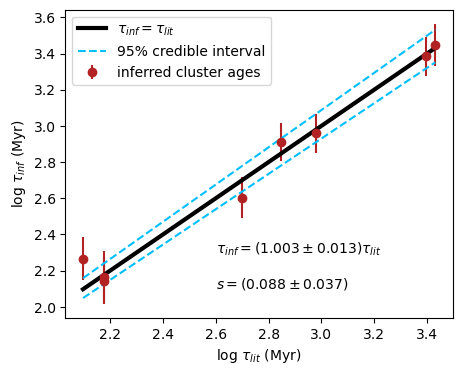}}
\caption{A summary of the cross-validation results by cluster, displaying the mean cluster posterior ages inferred by our framework ($\tau_{inf}$) against the literature ages ($\tau_{lit}$).}
\label{fig: loocv}
\end{center}
\vskip -0.2in
\end{figure}

Figure \ref{fig: loocv} summarizes the accuracy of the inferred $p(\tau_{cl})$ distributions from our LOOCV exercise for each cluster relative to literature. The best fit value and 95\% credible interval for $\theta$ from eq. \ref{eq: mcmc} are also highlighted. The $\tau_{inf}$ : $\tau_{lit}$ relationship is consistent with 1, indicating that the overall inference framework does not appear to suffer from systematic bias. One notable outlier here is our estimation for the Pleiades cluster (dated by Godoy-Rivera et al. \yrcite{2021ApJS..257...46G} as 125 Myr); this is unsurprising as \textit{(i)} it is the youngest cluster we use so there is no lower bound to interpolate against in our LOOCV exercise, and \textit{(ii)} the faster evolution of $P_{rot}$ in younger stars requires more thorough calibration to enable precise inference for lower ages. More robust uncertainty handling will be addressed in future work, however it is encouraging that the recovered $p(\tau_{cl})$ distributions, particularly when intrinsic scatter is included, are reasonable.

\section{Discussion} \label{sec: discussion}

It is important to consider these results alongside the precision of other stellar age dating techniques, and within the context of historical performance of analytical and empirical gyrochronology models. The overall uncertainties on our inferred cluster ages are dominated by the intrinsic scatter, and range from 0.088 to 0.99 dex. These are consistent with isochrone fitting uncertainties, which are typically in the 10-20\% range \cite{2021ApJS..257...46G}. The Gaia Collaboration et al. \yrcite{2018A&A...616A..10G} cite [-0.06,0.08] dex uncertainties for cluster ages greater than 100 Myr. The literature ages of NGC 6819 and Ruprecht 147 \cite{2020ApJ...904..140C} come with uncertainties of $\approx$ 0.03 dex based on a gyrochronological estimate of NGC 6819 calibrated to an isochrone model of Ruprecht 147. In a direct comparison of isochrone-only modeling to a combination of isochrone fitting and gyrochronology, Angus et al. \yrcite{2019AJ....158..173A} reported uncertainties of 22\% using only isochrones and 8\% with both techniques. The results of these studies indicate that our precision is generally comparable to current stellar dating standards.

Building off of these results, and given that we still consider this work to be in the proof of concept stage, there are several features of our model that we intend to improve. First, further exploration and optimization of the NF parameters, using both empirical evaluation and more prescriptive approaches to tuning, are crucial to improving performance. Second, we need to account for observational uncertainties and selection biases; ensuring that we correctly propagate these is essential to realistically assessing the credibility of our model and systematic errors that may be introduced. In particular, a more robust method of calculating $p(\tau_{cl})$ is required to generate proper statistical uncertainties. Finally, a deeper investigation into a wider parameter space for our model could dramatically improve performance (eg. metallicity, a factor linked to stellar rotational evolution \cite{2023arXiv230701688S}).

There are further open questions that we will resolve through additional testing and validation. M50 and NGC 2516 (both 150 Myr) are the only coeval clusters in our observational data, so the effect of including more needs to be explored. This is particularly true in the case of young clusters, where the complexity of rotational evolution will require more extensive calibration. Furthermore, the LOOCV exercise demonstrated that our NF performed well within this POC, however an entire cluster's data was excluded from training during each iteration. Applying the same evaluation framework to subsets and mixtures of cluster data will provide a more nuanced assessment of how our model will behave once we incorporate new stellar observational data in which cluster membership is less clear. Finally, developing more thoughtfully motivated priors for our stellar age and conditional colour distributions will provide a more realistic indication of our best-case performance given the current model.

\section{Conclusion} \label{sec: conclusion}

Our data-driven probabilistic model has successfully identified gyrochonological patterns in observational data to a precision comparable with empirical gyrochronological models and other stellar dating techniques. A particularly notable result is the successful interpolation of $p(P_{rot}\;|\;\tau\!,\!C)$, in the context where significant (age) gaps exist in the parameter space of training data. The efficacy of our framework will increase as more photometric data become available over time, but to properly leverage those observations it will be important to improve our model's ability to handle uncertainties and observational bias. With the combination of these enhancements, additional training data, and an extensive validation pipeline, we are optimistic that this framework can become a valuable tool in the field of stellar dating.

\bibliography{refs}
\bibliographystyle{icml2023}

%%%%%%%%%%%%%%%%%%%%%%%%%%%%%%%%%%%%%%%%%%%%%%%%%%%%%%%%%%%%%%%%%%%%%%%%%%%%%%%
%%%%%%%%%%%%%%%%%%%%%%%%%%%%%%%%%%%%%%%%%%%%%%%%%%%%%%%%%%%%%%%%%%%%%%%%%%%%%%%
% APPENDIX
%%%%%%%%%%%%%%%%%%%%%%%%%%%%%%%%%%%%%%%%%%%%%%%%%%%%%%%%%%%%%%%%%%%%%%%%%%%%%%%
%%%%%%%%%%%%%%%%%%%%%%%%%%%%%%%%%%%%%%%%%%%%%%%%%%%%%%%%%%%%%%%%%%%%%%%%%%%%%%%
% \newpage
\appendix
\onecolumn

\begin{appendices}
\begin{center}
\vspace{40mm}
\Large \textbf{APPENDIX}
\end{center}
\section{Cluster data details} \label{sec: app_clustes}

Table \ref{tab: clusters} provides the breakdown of observational data from open clusters used in this work.

% \begin{deluxetable}{ccrr} \label{tab:clusters}
% \tablewidth{210px}
% \tablehead{\colhead{Cluster} & \colhead{Source} & \colhead{\# Stars} & \colhead{Age (Myr)}}
% \tablecaption{Open cluster data used. The age of each cluster is defined as per its respective source. Sources are [1] \cite{2021ApJS..257...46G} and [2] \cite{2020ApJ...904..140C}.}
% \startdata
% Pleiades & [1] & 960 & 125  \\
% M50 & [1] & 131 & 150  \\
% NGC 2516 & [1] & 316 & 150 \\
% M37 & [1] & 213 & 500 \\
% Praesepe & [1] & 1032 & 700 \\
% NGC 6811 & [1] & 161 & 950 \\
% NGC 6819 & [2] & 30 & 2500 \\
% Ruprecht 147 & [2] & 35 & 2700
% \enddata
% \end{deluxetable}

\begin{table}[H]
    \centering
        \caption{Open cluster data used. The age of each cluster is the literature age as cited per its respective source: [1] \cite{2021ApJS..257...46G} and [2] \cite{2020ApJ...904..140C}.}
        \vspace{10mm}
    \begin{tabular}{||c | c | c| c||} 
          \hline
 Cluster & Source & \# Stars & Age (Myr) \\ [0.5ex] 
 \hline\hline
 Pleiades & [1] & 960 & 125  \\
  \hline
  M50 & [1] & 131 & 150  \\
  \hline
NGC 2516 & [1] & 316 & 150   \\
  \hline
M37 & [1] & 213 & 500 \\
  \hline
Praesepe & [1] & 1032 & 700 \\
  \hline
NGC 6811 & [1] & 161 & 950 \\
 \hline
NGC 6819 & [2] & 30 & 2500 \\
 \hline
Ruprecht 147 & [2] & 35 & 2700 \\
\hline 
    \end{tabular}
    \label{tab: clusters}
\end{table}
%%%%%%%%%%%%%%%%%%%%%%%%%%%%%%%%%%%%%%%%%%%%%%%%%%%%%%%%%%%%%%%%%%%%%%%%%%%%%%%
%%%%%%%%%%%%%%%%%%%%%%%%%%%%%%%%%%%%%%%%%%%%%%%%%%%%%%%%%%%%%%%%%%%%%%%%%%%%%%%
% % \pagebreak
\vspace{100px}
\section{All LOOCV results} \label{sec: app_all_loocv_res}

Here we present the naive $p(\tau_{cl})$ distributions from our LOOCV exercise for each cluster, both with and without intrinsic scatter, analogous to panel \textbf{e)} in figure \ref{fig: post_flow}.

\begin{figure*}[b]
    \centering
    \includegraphics[width=\textwidth]{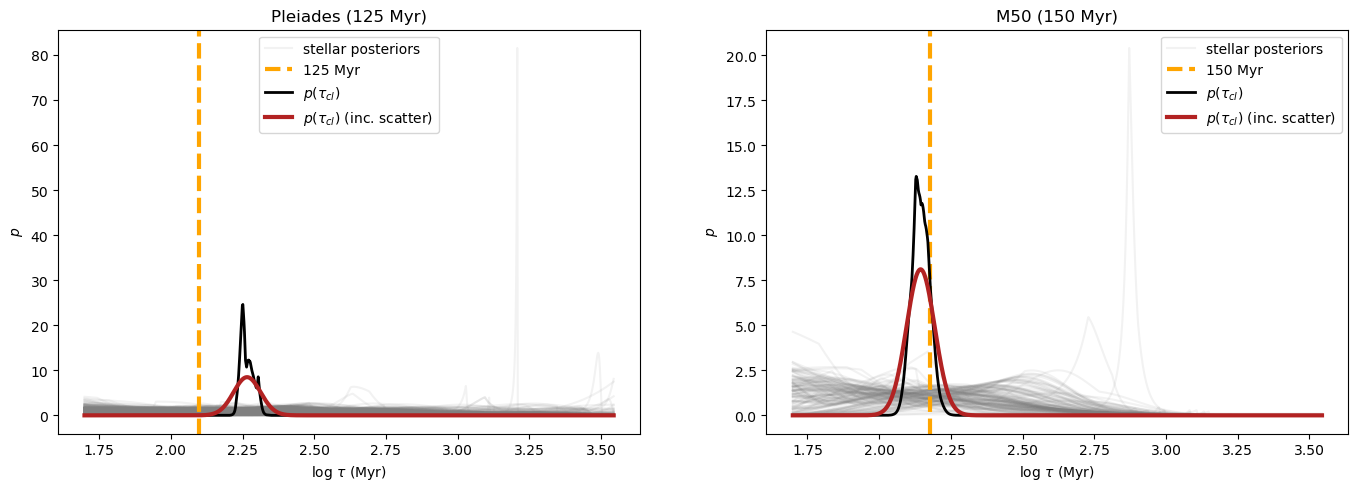}\\
    \label{fig: cluster_probs_all}
\end{figure*}

\begin{figure*}
    \vspace{10mm}
    \includegraphics[width=\textwidth]{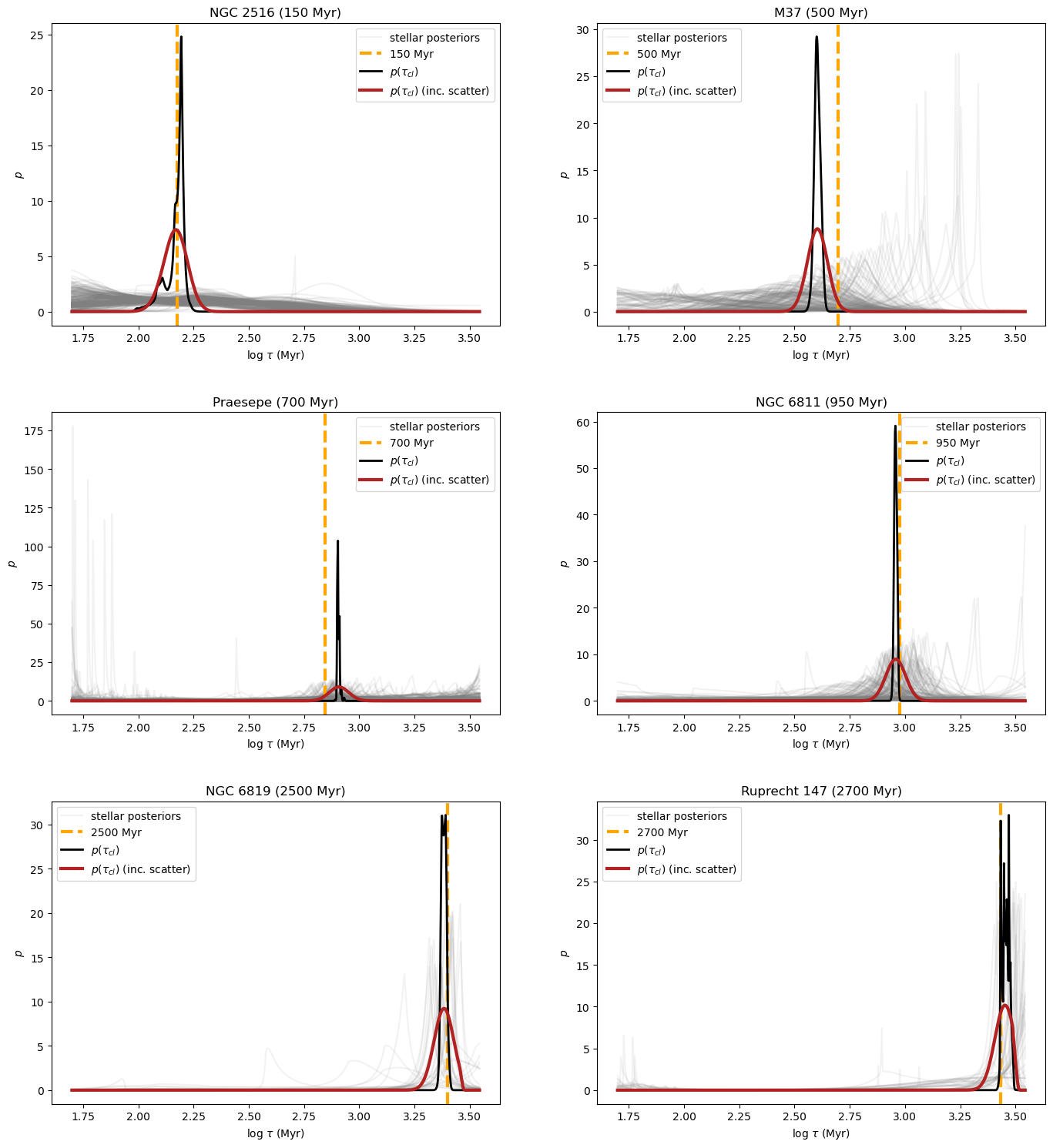}
    \caption{Plots of the naive $p(\tau_{cl})$ distributions, both with (red) and without (black) the fitted intrinsic scatter for each cluster. Also shown are all of the individual stellar posteriors (grey) and the literature cluster age (dashed orange line). Results are all from the NFs trained during our LOOCV exercise.}
\end{figure*}

%TC:endignore

\end{appendices}
\end{document}